\documentclass[12pt,a4paper]{article}

 \usepackage[dvips]{graphics}
 \usepackage{graphicx}
 \usepackage{afterpage}
 \usepackage{enumerate}
\begin{document}

 \title{\bf Horizontal Magnetic Fields in the Solar Photosphere}

 \author{\bf V.A. Sheminova}
 \date{}

 \maketitle
 \thanks{}
\begin{center}
{Main Astronomical Observatory, National Academy of Sciences of
Ukraine
\\ 27 Akademika Zabolotnogo st., Kyiv, 03680 Ukraine\\ E-mail: shem@mao.kiev.ua}
\end{center}

 \begin{abstract}
The results of 2D MHD simulations of solar magnetogranulation are
used to analyze the horizontal magnetic fields and the response of
the synthesized Stokes profiles of the FeI $\lambda$1564.85 nm
line to the magnetic fields. Selected 1.5-h series of the 2D MHD
models  reproduces a region of the network fields with their
immediate surrounding on the solar surface with  the unsigned
magnetic flux density of 192~G. According to the magnetic field
distribution obtained, the most probable absolute strength of the
horizontal magnetic field at an optical depth of $\tau_5$ = 1
($\tau_5$ denotes $\tau$ at $\lambda$ = 500 nm) is 50 G, while the
mean value is 244 G. On average, the horizontal magnetic fields
are stronger than the vertical fields to heights of about 400 km
in the photosphere due to their higher density and the larger area
they occupy. The maximum factor by which the horizontal fields are
greater is 1.5. Strong horizontal magnetic flux tubes emerge at
the surface as spots with field strengths of more than 500 G.
These are smaller than granules in size, and have lifetimes of 3.6
min. They form in the photosphere due to the expulsion of magnetic
fields by convective flows coming from deep subphotospheric
layers. The data obtained qualitatively agree with observations
with the Hinode space observatory.
\end{abstract}

\section{Introduction}

Recent spectropolarimetric observations using the Hinode satellite
\cite{1} have revealed unexpected properties of the magnetic
fields in the quiet Sun. Horizontal fields proved to have larger
magnetic flux densities and occupy larger areas on the solar
surface than vertical fields \cite{2, 3}. Information in the
literature about the properties of horizontal magnetic fields in
the solar photosphere is relatively sparse. Such fields were first
observed in inter-network photospheric regions as short-duration
(5 min) phenomena related to the decay of emerging weak bipolar
magnetic fluxes \cite{4}. The results of a Stokes-profile
inversion \cite{5} confirmed the presence of horizontal fields and
demonstrated that weak horizontal fields should be present in
quiet regions along with relatively strong vertical fields. Later,
horizontal fields were detected in low-lying photospheric loops
(bipoles) \cite{9,7,6,8} and in unipolar loops of emerging
magnetic flux \cite{10}. Magnetographic full disk observations of
the Sun  \cite{11} also detected many horizontal fields in the
quiet photosphere and their time variations. In addition, isolated
small-scale islands of strong horizontal fields were recently
observed in facular areas (plages) \cite{12}.

The currently available observational results lead to the
following conclusions about the properties of horizontal magnetic
fields in the solar photosphere.

(1) Horizontal fields are ubiquitous at the surface of the quiet
Sun. They occupy a larger area than vertical fields, and are
located nearer the edges of granules and intergranular lanes.

(2) The structures in the horizontal fields are typically somewhat
smaller than granules and larger than structures in the vertical
fields.

(3) The horizontal fields are spatially separated from the
vertical fields.

(4) The flux density of the horizontal component of the magnetic
field in the inter-network regions (55 G) is a factor of five
higher than the flux density of the vertical component (11 G).

(5) Strong horizontal fields in plages appear as small islands
with strengths reaching 580 G.

What is the source of the abundant horizontal fields at the solar
surface? How are the horizontal fields related to the vertical
fields? What is the height variation of the horizontal fields in
the photosphere? How are they distributed over the surface? How
strongly are they coupled with penetrative convection processes?
What is the nature of the horizontal photospheric fields? These
questions can be addressed by magnetohydrodynamical (MHD)
simulations of photospheric regions. No detailed analyses of the
properties of the horizontal fields based on MHD simulations has
been done \cite{14,15,13,16}, and interest in these fields has
been revived with the availability of Hinode observations. New
three-dimensional (3D) simulations of inter-network fields in the
quiet photosphere were carried out recently in \cite{17}.
According to the results obtained for a level of mean optical
depth, $\tau$ = 1, horizontal fields with strengths of $>$50~G
cover an area a factor of three larger than the area covered by
vertical fields with the same strengths. A local maximum of the
horizontal field component is located near the temperature
minimum($\approx 500$~km). The strength of the horizontal field
component exceeds that of the vertical component by factors of 2.0
or 5.6, depending on the initial conditions of the simulations.
The horizontal fields are strongly coupled with penetrative
convection.

Three-dimensional MHD simulations of local dynamo processes in
subsurface solar layers were carried out in the subsequent study
\cite{18}. An analysis of the magnetic fields demonstrated that
the horizontal field component dominates the vertical component in
the height range where the photospheric spectral lines form. The
ratio of the two components corresponded to the observed value.
Magnetic fields of mixed polarities produced by a local dynamo
mechanism near the surface are the main source of horizontal
fields at the surface of the quiet Sun. Analyzing some other
possible mechanisms \cite{18}, it was also noted that the
formation of horizontal fields in the photospheric network and
plages is affected more by the recirculation of the overall
background granulation flux than by dynamo processes.

Our aim here is to analyze photospheric horizontal fields in the
region of the photospheric magnetic network using two-dimensional
(2D)MHD models made by Gadun \cite{24,14,15}, in order to answer some questions
concerning their properties and nature.

In Section 2, we briefly describe the 2D MHD models, our analysis
of the horizontal fields based on data of simulated
magnetoconvection is presented in Section 3, the synthesized
Stokes profiles are analyzed in Section 4, Section 5 presents
comparisons with observations and a brief discussion, and Section
6 summarizes our conclusion.

\section{MHD models of magnetogranulation}

Realistic MHD models of photospheric regions are self-consistent
models obtained by solving a system of radiative MHD equations.
They can be used to synthesize spectral lines without any
additional parameters such as the microturbulent and
macroturbulent velocities. A review of the most recent results of
MHD simulations can be found in \cite{19}. Two-dimensional
simulations of the solar convection have been actively developed
in parallel with 3D simulations since 1984. Although 2D models
cannot realistically reproduce convective flows \cite{20}, they
reproduce many characteristics of 3D convection quite adequatly
\cite{21, 22} and are useful in studying the properties of
small-scale magnetic elements in the solar photosphere
\cite{23,24, 15,13,25}.

Among the known 2D MHD models for the solar magnetogranulation
\cite{ 23,27, 26,14,13},  Gadun's models \cite{24,14,15}
have proven to be fairly successful. They
reproduced for the first time the small-scale magnetic fields of
mixed polarities outside the active regions in the solar
photosphere (see for details \cite{24}). Based on these models, a
close relationship between the strength and inclination angle of
the photospheric magnetic fields and the predominance of
horizontal fields were inferred \cite{15}. This agrees with
observations \cite{6} obtained using the Advanced Stokes
Polarimeter (ASP). Further, first evidence for the recirculation
of convective flows near the surface was found \cite{28}, and
later confirmed by 3D MHD simulations \cite{29}. Using the 2D MHD models
for Stokes diagnostics predicted specific
changes in the Stokes profiles of the FeI $\lambda$ 1564.8 nm line
during the convective collapse of magnetic tubes \cite{25}, which
was soon confirmed by observations with the Tenerife Infrared
Polarimeter (TIP) \cite{30}. Gadun's  models  made
it possible to explain the observed anomalous asymmetry of the
Stokes profiles of the FeI $\lambda$630.2 nm line \cite{31, 32}
and investigate the distribution of the vertical magnetic-field
component in regions with varying magnetic-flux densities at the
solar surface, outside active regions  \cite{33}.

The active development of new, improved 3D MHD models for the
solar magnetoconvection is currently underway  \cite{16, 17, 34,
35}; however, they cannot yet answer all of the numerous questions
that arise when interpreting the observational results \cite{19}.
Therefore, using the 2D MHD magnetogranulation models to
study the small-scale structure of magnetic fields remains
important.

The series of 2D MHD models are used in this paper  was obtained
by solving a full system of radiative MHD equations for a
compressible, gravitationally stratified, turbulent medium \cite{14}. Free
boundary conditions admitting material inflow or outflow were
specified for the velocities and thermodynamic quantities at the
top and bottom boundaries. The corresponding boundary conditions
for the magnetic field were $B_{hor} = 0$,
$\partial{B_{ver}}/\partial{ z} = 0$. The side boundary conditions
were assumed to be periodic, which corresponds to multiple mirror
reflections of the simulation domain in both horizontal
directions. The simulation domain was located at heights ranging
from 685 km above to 1135 km below the surface level, $Z = 0$~km.
The horizontal size of the domain was about 4000~km, which
corresponds to $5.5^{\prime\prime}$ on the solar disk. The spatial
step was 35 km. The MHD simulations started by introducing a
bipolar magnetic field in a previously computed two-dimensional
hydrodynamic (2D HD) thermal-convection model. The initial
unsigned magnetic flux was 54 G throughout the simulation
domain. Within 30~min after the beginning of the simulations, the
magnetic field comes into consistency with the convective motions
of the plasma. Therefore, for further analyses, sampled MHD models
can be used starting from a time of 30 min. More details of the
numerical 2D MHD simulations can be found in  \cite{24,14, 15}.

The mean magnitudes of the vertical and horizontal components of
the simulated magnetic field at an optical depth at $\lambda$ 500
nm of $\tau_5$ = 1 grow rapidly with time, and fluctuate
appreciably (Fig. 1). The flux density of the horizontal field at
the surface increases due to the expulsion of field from deep
layers by convective flows and its accumulation in the upper
layers of granules. Further, the flux densities of the horizontal
fields decrease sharply, since the fields flow into intergranular
lanes and become nearly vertical. In narrow intergranular lanes,
convective collapse comes into action after the vertical fields
reach their equipartition level, and can either enhance the
vertical fields in magnetic tubes to 2000 G or completely destroy
these tubes. A detailed analysis of magnetoconvection  \cite{24}
demonstrates that the flux density increases and oscillations in
the magnetic flux are due to repeated penetrative-convection
processes. One specific example of local recirculation of the
granulation flux in the considered 2D MHD models is discussed in
detail in  \cite{28}.
\begin{figure}
\centerline{\includegraphics[width=0.55\textwidth,clip=]{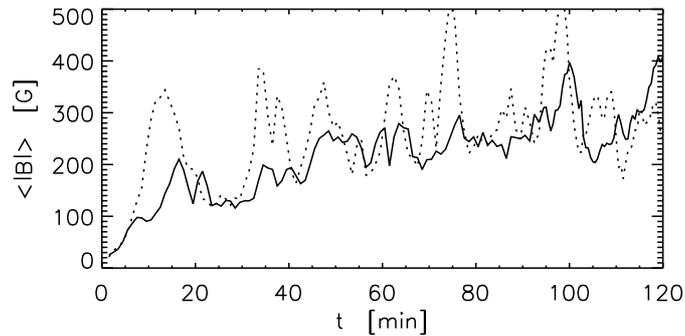}}
  \caption{Simulated mean magnetic-field
strength at the level $\tau_5 = 1$ as a function of time in the
series of 2D MHD models. The solid curve shows $<|B_{ver}|>$ and
the dashed curve $<|B_{hor}|>$. }
\end{figure}

\begin{figure}[t!]
\centerline{\includegraphics[width=0.95\textwidth,clip=]{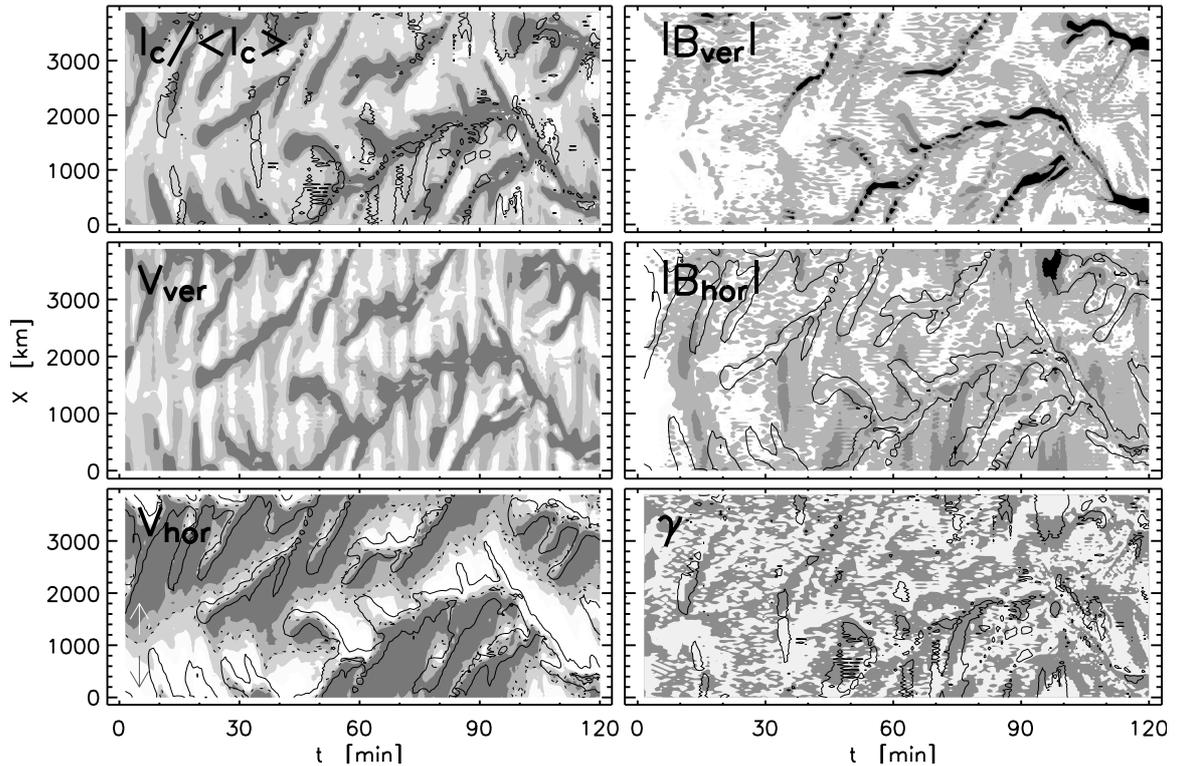}}
  \caption{Spatial and temporal evolution of
the simulated 2D magnetogranulation at the continuum-formation
level, $\tau_5$ = 1. The top left panel shows the relative
intensity of the continuum at a wavelength of 500 nm, with the
gray scale from dark to light corresponding to $I_c/<I_c>$~values
of $<$0.8, 0.8--1, 1--1.3, and $>$1.3. The contours indicate
islands of horizontal fields with $500$~G~$<|B_{hor}|<1000$~G. The
middle left panel shows the line-of-sight velocity $V_{ver}$ with
the gray scale from dark (descending flows) to light (ascending
flows) corresponding to $V_{ver}$  values of $<$-1.5, -1.5--0,
0--1.5, and $>$1.5~km/s. The bottom left panel shows the
horizontal velocity $V_{hor}$; the gray scale is similar to that
for $V_{ver}$. The arrows near the lower left corner indicate the
directions of the matter flow. The contours mark regions of
intergranular lanes, and the dashed curve corresponds to
$V_{hor}=0$. The top right panel shows the magnitude of the
vertical component, with the gray scale from light to dark
corresponding to $|B_{ver}|$ values of~$<$100, 100--500,
500--1000, and $>$1000~G.The middle right panel shows the
magnitude of the horizontal component; the gray scale is similar
to that for $|B_{ver}|$.The contours mark regions of intergranular
lanes. The bottom right panel shows the inclination angle,
$\gamma$, of the magnetic field; light gray shows positive and
dark gray negative values. The contours indicate islands of
horizontal fields.}
\end{figure}

Figure 2 shows the spatial and temporal variations in the magnetic
field, velocity field, and radiation field at the $\tau_5$ = 1
level for the series of 2D MHD magnetogranulation models. The
spatial-temporal image of the continuum intensity, $I_c$,
represents the evolution of the granulation. Light regions
(granules) can be seen between dark, twisted strips (intergranular
lanes). The fragmentation of the granules, i.e., the separation of
large granules into two smaller ones, is pronounced. We can
clearly see that the horizontal size of the granules decreases as
the magnetic flux increases in the simulation domain. The vertical
velocity field, $V_{ver}$, closely resembles the corresponding
diagram for the continuum intensity, since the radiation intensity
and the line-of-sight velocity at the solar surface are strongly
correlated. Note that the 5-minute oscillations of the
line-of-sight velocity are more pronounced than the intensity
oscillations. To trace the behavior of horizontal matter flows and
their relationship to the granulation, we also show contours of
the intergranular lanes in the plot of the horizontal velocities
$V_{hor}$. The matter spreads from the centers to the edges of the
granules. The horizontalflow velocities are substantially enhanced
at the edges of magnetic tubes located in the intergranular lanes.

The plots of the vertical, $B_{ver}$, and horizontal, $B_{hor}$,
components of the magnetic field shown to the right in Fig. 2 are
shown using the same scale for the field strengths, to make these
values and the areas occupied by them easily comparable. There are
substantial differences between the vertical and horizontal
fields, in both the shapes of structures and magnetic flux
densities. Islands of strong horizontal fields are present in
granules, mostly relatively close to intergranular lanes. As a
rule, islands located at the centers of granules emerge before the
fragmentation of the granule and the formation of new magnetic
tubes. We described the processes of granular fragmentation in
detail earlier in  \cite{14, 15}. The plot of the magnetic-field
polarity, $\gamma$, demonstrates strong small-scale mixing of
magnetic fields directed toward and from the observer.

Strong vertical magnetic concentrations are clearly visible after
the first 30 min of the simulations in the top right panel of Fig.
2. Their outlines resemble the observed magnetic structures of the
photospheric supergranular network, which are surrounded by
weaker, mixed fields. In addition, the flux density of the
vertical magnetic field at the $\tau_5 = 1$ level in the region of
the simulated magnetogranulation is close to the value observed in
the vicinity of the photospheric network  \cite{36}; i.e., it is
an order of magnitude higher than the flux density in
inter-network areas. The amplitudes of the Stokes profiles of
spectral lines computed for the given simulation domain are also
an order of magnitude larger than the amplitudes measured in
inter-network regions. We thus conclude that our simulations have
reproduced the emergence of magnetic flux through the photosphere
in network rather than inter-network areas.

Our analysis is based on a 1.5-h series of 2D MHD models (starting
from the first 30 min of the simulations). The time resolution is
1 min during the first hour and 0.5 min during the subsequent 30
min. This series forms 126 2D models, which, in turn, contain 112
1D models each. In total, we have 14 112 magnetic-field
measurements, enabling a statistical analysis of the simulation
data with the aim of revealing the properties of the horizontal
fields. We emphasize that, according to  \cite{37}, the spatial
resolution of the numerical MHD simulations corresponds to two or
three steps of the computational grid. For our series of 2D MHD
models, the resolution is 100 km, or $0.15^{\prime\prime}$.

\section{Properties of the horizontal magnetic fields}

Figure 3 shows the strengths of the vertical, $|B_{ver}|$, and
horizontal, $|B_{hor}|$, components averaged over space and time
for the 1.5-h series of 2DMHD models as a function of optical
depth. The mean strength of the horizontal field in the
photosphere is higher than the mean strength of the vertical field
down to the level $\log \tau_5\approx-3.5$ ($Z\approx400$~km). The
maximum excess, $<|B_{hor}|>/<|B_{ver}|> \approx 1.5$, is
substantially less than is observed for the inter-network areas of
the quiet Sun  \cite{3}.
\begin{figure}
\centerline{\includegraphics[width=0.55\textwidth,clip=]{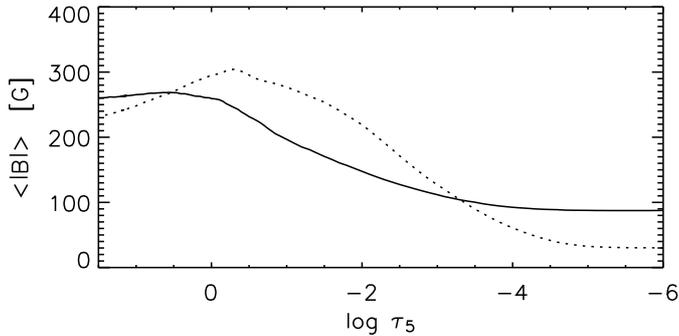}}
\caption{ Mean magnitudes of the horizontal and vertical
components of the magnetic field in the 1.5-h series of 2D MHD
models as a function of the height in the photosphere. The solid
curve shows $<|B_{ver}|>$, and the dashed curve $<|B_{hor}|>$. }
\end {figure}

The probability-density function (PDF) is frequently used to
investigate the distribution of the magnetic-field strength over
the surface  \cite{38}. Figure 4 shows the PDFs obtained
for $|B_{ver}|$, $|B_{hor}|$, and the inclination of the
magnetic-field vector, $\gamma$, at the $\tau_5$ = 1 level. The
most probable values for $|B_{ver}|$, $|B_{hor}|$, and $\gamma$
are 50~G, 50~G, and 87$^\circ$, and the corresponding mean values
are 192~G, 244~G, and 64$^\circ$, respectively. Hence, the
factor by which the horizontal exceeds the vertical field
component averaged over the domain at the $\tau_5$ = 1 level is
1.3.

\begin{table}[!]
\centering
\caption{Percentages of weak, moderate, and strong magnetic fields
in the 1.5-h series of 2D MHD models in terms of their occupied
areas on the surface of the simulation domain ($\alpha$), their
densities ($\phi$), and the magnetic energy they produce
($\varepsilon$).
 }
\vspace{0.3cm}
\begin{tabular}{|c|c|c|c|}
\hline

 Strength, G    & $\alpha~(\%)$& $\phi~(\%)$&$\varepsilon~(\%)$\\
\hline
 $|B_{hor}|<500$ &          72          &   28       &
 15\\
 $500\leq|B_{hor}|<1000$ &  18          &   20       &
 30\\
 $|B_{hor}|\geq 1000$ &     10          &   2        & 5\\
\hline
 $|B_{ver}|<500$ &         67          &   25       & 8\\
 $500\leq|B_{ver}|<1000$ &   9          &   12       &
 12\\
 $|B_{ver}|\geq 1000$ &     24          &   13       &
 30\\
[1mm]
  \hline
\end{tabular}
\end{table}

Following  \cite{38}, we calculate the filling factor $\alpha$
based on the obtained PDFs, which determines the fractions of the
surface occupied by weak, moderate, and strong fields, which we
arbitrarily define as fields with strengths $|B_{hor}|< 500$~G,
$500\leq|B_{hor}|< 1000$~G, and $|B_{hor}|\geq 1000$~G. We apply
similar definitions to the vertical fields. Note that, in contrast
to the observed patterns, the surface of the simulation domain is
completely covered with magnetic field, i.e., the magnetic-field
filling factor is unity. In addition to $\alpha$, we also
calculated the fractions of weak, moderate, and strong fields in
terms of their magnetic fluxes $\phi$ and the fractions of the
magnetic energy associated with them relative to the total
magnetic energy $\varepsilon$. The results of our calculations in
the total magnetic energy are given in the table. These indicate
that weak horizontal fields in the simulated region of the
photospheric network occupy a larger area (72\%) and have a higher
flux density (28\%) than the other fields. Relatively large
contributions are given by moderate horizontal and strong vertical
fields (30\% each). The moderate horizontal fields can be seen as
small islands in Fig. 2 (second plot on the right). These are
mainly bipolar, and their lifetimes are 3-6 min. In many cases,
such islands emerge in regions next to intergranular lanes rather
than in the central parts of the granules.
\begin{figure}[!]
\centerline{\includegraphics[width=0.95\textwidth,clip=]{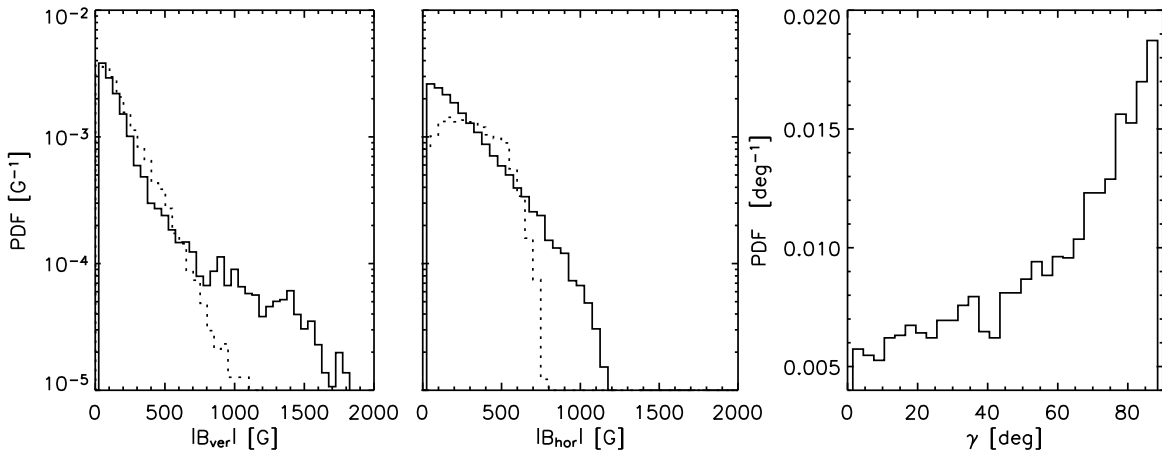}}
  \caption{ Histograms of the
probability-density functions (PDFs) for the strengths of the
vertical and horizontal components of the magnetic field and the
inclinations of the magnetic vector in the 1.5-h series of MHD
models at the $\tau_5$ = 1 level. The solid curve shows the
MHD-simulation data, and the dotted curve the Stokes-diagnostic
data. }
\end {figure}

Strong horizontal fields cover 10\% of the total surface area,
while strong vertical fields cover 24\%. An island of strong
horizontal field is clearly visible in Fig. 2 (second plot╩ on the
right) in a large granule, shortly before its fragmentation. In
contrast to strong horizontal fields (or horizontal tubes), strong
vertical fields (or vertical tubes) are concentrated in
intergranular lanes (Fig. 2, first plot on the right). These
fields form a network of strong vertical fields in the simulation
domain, similar to the observed fields of the photospheric network
represented by magnetograms.
\begin{figure}
\centerline{\includegraphics[width=0.4\textwidth,clip=]{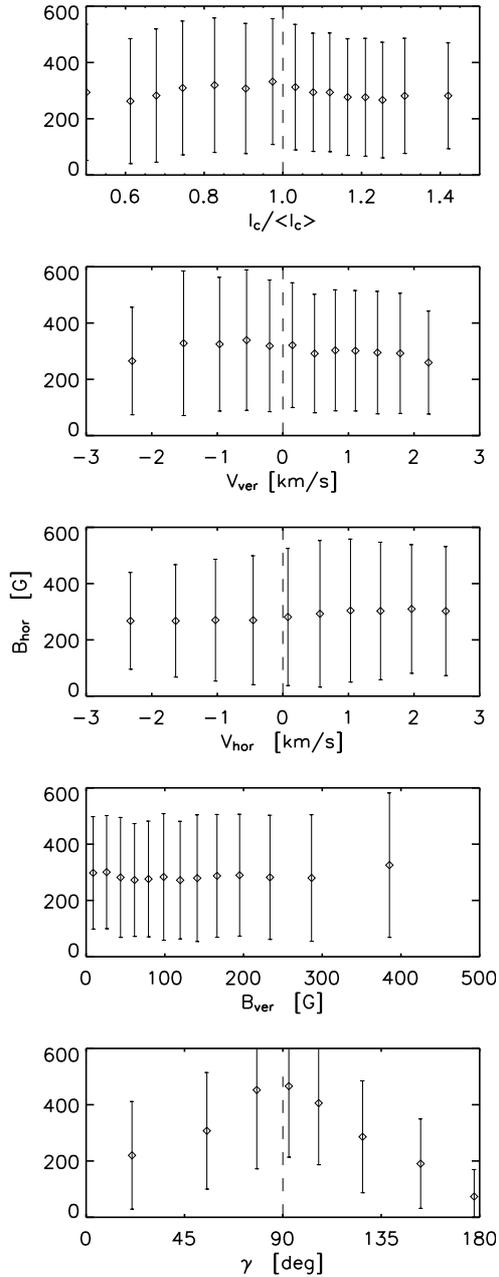}}
 \hfill
 \caption{ Statistical dependences of the
strengths of the horizontal magnetic-field component on the
parameters characterizing the granulation and magnetic field.
Positive $V_{ver}$ correspond to upflows and
negative $V_{ver}$ to downflows. The values are averaged over the
same number of points (881) in each interval. The vertical bars
indicate the standard deviation (the spread of values) in each
averaging interval. The rms error of each averaged value of
$B_{hor}$ is typically $<$10 G. }
\end {figure}

Let us consider the properties of the horizontal fields in the
simulation domain using the statistical dependences of $B_{hor}$
on the granulation parameters (Fig. 5). The relationship of the
horizontal fields to the granulation reflects the dependence of
$B_{hor}$ on the contrast of the continuum emission, $I_c/<I_c>$.
In the figure, $I_c$ is the continuum intensity at $\lambda$ 500
nm at each computation point, and $<I_c>$ is the intensity $I_c$
averaged over the simulation domain for a particular time. No
pronounced dependence can be seen, although stronger horizontal
fields exhibit some tendency to correspond to contrasts of
$I_c/<I_c>\approx1$ or somewhat less. The dependence of $B_{hor}$
on the line-of-sight velocity $V_{ver}$ is also not clearly
expressed, and there is only a weak tendency for $B_{hor}$ to
increase in granulation areas with slow, descending matter flows.
There is virtually no relationship between $B_{hor}$ and the
horizontal component of the velocity field. Similarly, $B_{hor}$
and $B_{ver}$ are uncorrelated. The dependence of $B_{hor}$ on the
inclination $\gamma$ indicates a strong relationship between them,
with the strength of the horizontal component being, on average,
about 500 G for the most inclined, nearly horizontal fields and
about 200 G for weakly inclined fields (on average,
$\gamma=20^\circ$). Thus, the statistical dependences do not
reveal any close relationships between the horizontal fields and
the granulation and vertical fields. The moderate and strong
horizontal and vertical fields are spatially separated.

\section{Stokes diagnostics of the magnetic fields}

To analyze the response of the linear and circular Zeeman
polarization to the magnetic fields, we considered synthesized
Stokes profiles ($I,~ Q,~U,~V$) of the FeI $\lambda$ 1564.8 nm
line. These profiles were obtained along the line of sight for the
1.5-h series of MHD models at any point of the simulated surface,
in accordance with observations at the center of the solar disk.
The spectral resolution was 5~m\AA~within $\pm$1.5~\AA~of the line
center. The iron abundance was chosen to be$A_{Fe}=7.43$, the
oscillator strengths $\log gf = -0.675$,, and the excitation
potential of the lower level $EP=5.43$~eV. The Van der Waals
damping constant was computed using the formulas given in
\cite{39}. The computations were carried out using the SPANSATM
code  \cite{40}. Analogous to the interpretation of observations
described in  \cite{3}, we obtained $V_{tot}$ and $Q_{tot}$ from
the synthesized Stokes $V$ and $Q$ profiles by integrating $V$ and
$Q$ over the wavelength:

\[V_{tot}=sgn(V_{b}) \frac{|\int^{\lambda_0}_{\lambda_b}V(\lambda)d\lambda|+
|\int^{\lambda_r}_{\lambda_0}V(\lambda)d\lambda|}
{I_c\int^{\lambda_r}_{\lambda_b}d\lambda }, \]

\[Q_{tot}=
\frac{\int^{\lambda_r}_{\lambda_b}|Q(\lambda)|d\lambda}
{I_c\int^{\lambda_r}_{\lambda_b}d\lambda }. \]

Here, $sgn(V_{b})$ is the sign of the blue peak amplitude of the
$V$ profile, $\lambda_0$ the central wavelength of the line, and
$\lambda_b$ and $\lambda_r$ the limits for integration over the
full line profile. Note that the intensity of the Stokes $U$
profile is nearly zero in 2D MHD simulations, and we did not
include it when calculating the total linear polarization.

To convert the computed $V_{tot}$ and $Q_{tot}$ values into field
strengths, we determined the dependences of $V_{tot}$ and
$Q_{tot}$ on $B_{ver}$ and $B_{hor}$. The strengths $B_{ver}$ and
$B_{hor}$ were deduced from 2D MHD models at the $\log \tau_5=-1$
level. According to our calculations, this is the photospheric
level that corresponds, on average, to the effective formation
depth of the Stokes profiles of the FeI $\lambda$ 1564.8 nm line.
The calibration curves are shown in Fig. 6. We used these to
obtain the magnetic flux density for the horizontal, $B_{hor}$,
and vertical, $B_{ver}$, components from each pair of synthesized
V and Q profiles and to construct the corresponding PDFs (Fig. 4,
dotted curve). These differ substantially from the PDFs obtained
from direct measurements of $B_{ver}$ and $B_{hor}$ using the MHD
models (solid curve). According to the PDFs based on the Stokes
profiles, the most probable strengths of the horizontal and
vertical magnetic-field components are 174 and 28 G, while the
mean values are 310 and 188 G, respectively. The factor by which
the mean horizontal exceeds the mean vertical field strength is
1.6. Recall that direct measurements using MHD models yielded 50
and 50 G, 244 and 192 G, and 1.3. The horizontal-excess factor
deduced from the Stokes profiles is slightly higher than the
factor obtained from the MHD data because the calibration curves
overestimated the flux density of the horizontal fields. This
means that the calibration of the linearpolarization signals
introduces an additional error in estimates of the strength of the
horizontal field.

According to the analysis of two series of 3D MHD models of
inter-network regions presented in  \cite{17}, the
horizontal-excess factors were 2.0 and 5.6 (MHD data) and 1.5 and
2.8 (Stokes diagnostics), i.e., the Stokes-based factors were
smaller than the simulated factors. In the analysis of  \cite{17},
the linear-polarization signals in the inter-network regions are
an order of magnitude weaker than in magnetic-network fields. This
could also affect the accuracy of the Stokes diagnostics and
calibration.

\begin{figure}
\centerline{\includegraphics[width=0.42\textwidth,clip=]{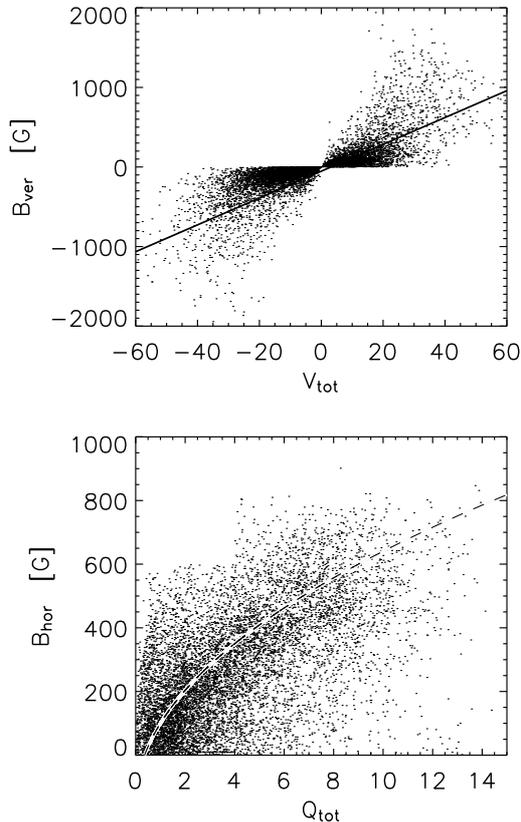}}
  \caption{ Calibration curves based on the
2D MHD simulations of $B_{ver}$ and $B_{hor}$ and the synthesized
Stokes profiles of the FeI $\lambda$1564.8 nm line for which the
circular ($V_{tot}$) and linear ($Q_{tot}$) polarizations were
calculated in m\AA.}
\end {figure}

\section{Comparisons with observations}

According to spectropolarimetry observations carried out on the
Hinode satellite with a spatial resolution of
$0.3^{\prime\prime}$, the flux density of the horizontal
magnetic-field component in inter-network regions is about 55~G
\cite{2, 3}, and reaches 580 G  \cite{12} in the plages. The mean
magnitude of the horizontal-field strength for our simulated
domain, which probably represents a region of the observed
supergranular photospheric network and its immediate vicinity, is
244 G. We believe that this is consistent with the observations.

Let us compare the available observational data for both
inter-network regions and the plages with the results of our
analysis. According to our 2D MHD simulations, the magnitude flux
density of the horizontal magnetic field exceeds that of the
vertical field by, on average, a factor of 1.3 (with the maximum
factor being 1.5). This is much less than the value derived from
observations of inter-network regions (a factor of 5  \cite{3}).
The smaller factor we have obtained in the simulated
photospheric-network region may be due to the higher density of
strong, small-scale, vertical network magnetic fields (magnetic
flux tubes) in the simulation domain, compared to the observed
inter-network quiet regions.

Observations of the plages  \cite{12} revealed the emergence of
isolated horizontal fields in the form of small islands
$1.4^{\prime\prime} \times 1 ^{\prime\prime}$ in size with
lifetimes of about 6 min. These appear inside a granule and
gradually move to the intergranular lanes, later leaving the field
of view. The inversion of the Stokes profiles observed in these
islands indicates that the horizontal field can reach 580 G.
According to our results, islands of strong horizontal fields with
similar sizes and lifetimes and with strengths of
$500~G$~$>B_{hor}>1000$~G also emerge inside granules at the
surface of the simulation domain (Fig. 2, second plot on the
right). In addition, we detected an island of strong field with
$B_{hor}>1000$~G, $X=3600$~km, and $t=95.5$~min.

Figure 7 shows a snapshot of the vertical cross section of the
simulation domain at the site where an island of strong horizontal
fields is forming. At time $t=95.5$~min, two strong horizontal
tubes and two strong vertical tubes can be seen. The convective
motions penetrating from lower subphotospheric layers carry
magnetic fields to middle and upper photospheric layers; this is
the effect of so-called magnetic flux expulsion, described in
detail in  \cite{18, 17}. A further growth of the field strength
in the horizontal tube that has formed on the surface of a granule
occurs under the action of the cool, denser matter located over
the large granule. The magnetic field lines condense and start to
bend. The cloud of cool material gradually sinks, entraining
magnetic-field lines. The large granule begins fragmenting, and
this pumps the strong horizontal field into deeper layers, forming
a new vertical tube  \cite{24,15}.
\begin{figure}[t!]
\centerline{\includegraphics[width=0.95\textwidth,clip=]{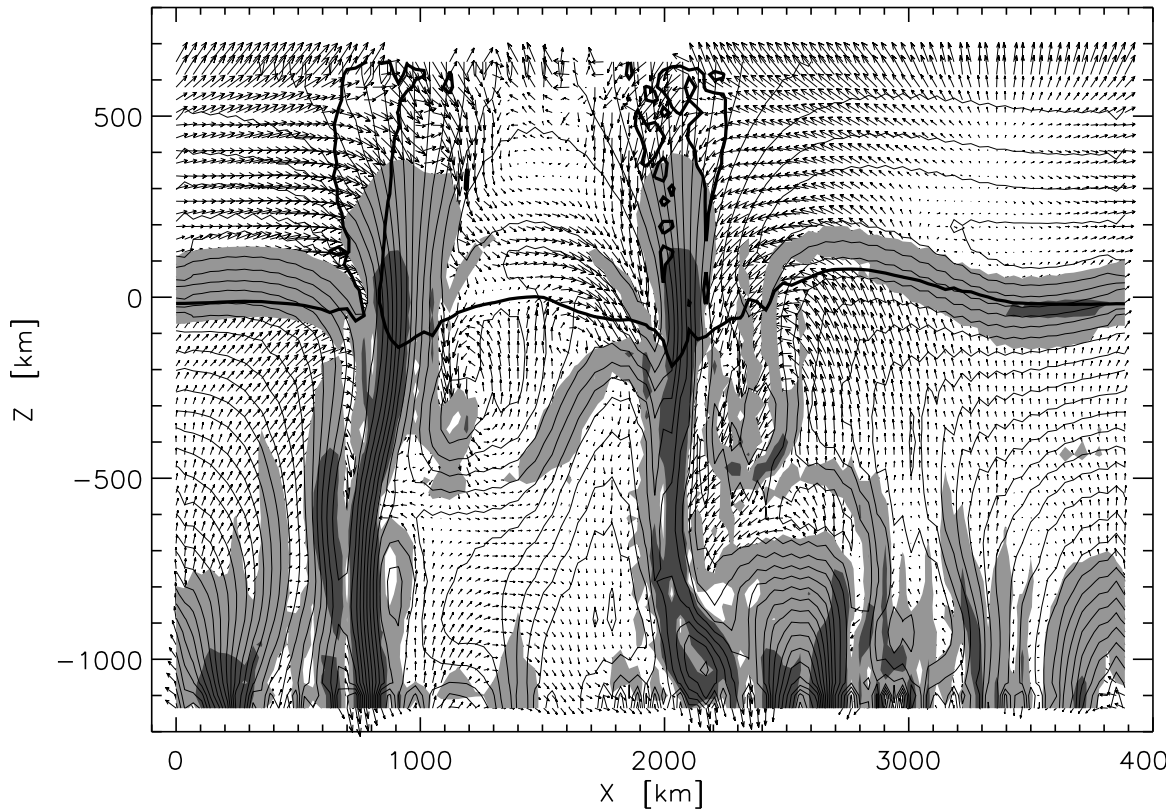}}
  \caption{  Snapshot of the vertical cross
section of the simulation domain for $t=95.5$~min. The heavy curve
shows the isotherm for T = 5500 K, which approximately indicates
the level of the mean optical depth of $\tau\approx 1$, while the
light solid curves show magnetic-field lines. The gray shading
shows vertical and horizontal fields with strengths from 500 to
1000 G, while dark gray corresponds to strengths above 1000 G. The
arrows indicate the direction of the velocity of the matter flow
at each computational grid point, and their length is proportional
to the corresponding flow speed. }
\end {figure}

It follows from our 2D MHD simulations that strong horizontal
tubes do not rise to upper photospheric layers (Fig. 7).
Therefore, the mean flux density excess of the horizontal over the
vertical component, shown in Fig. 1, is basically manifest in the
lower and middle photosphere. For the same reason, the area
occupied by strong horizontal tubes at the $\tau_5$ = 1 level far
exceeds the area of strong vertical tubes.

The response of the synthesized Stokes profiles of the iron
$\lambda$1564.8 nm line to horizontal and vertical fields in the
1.5-h series of MHD models yields a horizontal-excess factor of
1.6. This is a factor of three smaller than the value obtained
from observed Stokes profiles of the $\lambda$630.2 nm iron line
 \cite{3} in internetwork regions. The main reason for the disagreement
between the observations and our simulations appears to be the
difference in the magnetic fluxes for the observed and simulated
regions. On the other hand, our horizontal-excess factor is not
very different from the factors of 1.5 and 2.8 found in  \cite{17}
using the synthesized Stokes profiles of the FeI $\lambda$ 630.2
nm line and two series of 3D MHD models of internetwork regions.
The simulation results for internetwork fields of  \cite{17} do
not satisfactorily reproduce the factor of 5 obtained in \cite{3},
although the magnetic flux density in the 3D MHD models is similar
to the values derived from observations. One possible reason for
this disagreement is insufficient accuracy in the calibration of
the observed linear-polarization signals. We also cannot rule out
the possiblity that the vertical magnetic flux density was
underestimated in the observations of  \cite{3}, which had a
spatial resolution of  0.3$^{\prime\prime}$ (or about 200~km).
This could result if the contributions of smaller magnetic
structures with opposite field polarities partially cancel out,
which was not taken into account in the estimates of the flux
density.

Note also that the 2D MHD simulations of magnetoconvection that we
use here have some limitations and drawbacks. The 2D
representation of the granular motions is a very crude assumption.
The matter-flow divergence in the 2D simulation domain occurs in a
plane, which can affect the velocities and possibly the
magnetic-field strengths. The magnetic tubes in our models form
between two granules, while small-scale elements in the solar
photosphere are observed in gaps joining three or more granules.
The simulations deal only with the granular scale of the solar
convection, and the initial magnetic field is artificially
introduced in the models. To comprehend and trace the formation of
magnetic-network regions, it would be desirable to take into
account larger scales of the solar convection, such as the meso-
and supergranulation scales. These drawbacks of the simulations
could affect the quantitative estimates obtained. Thus, our
conclusions below are based on the qualitative features of our
results.

\section{Conclusion}

We have analyzed horizontal magnetic fields using a 1.5-h series
of 2D MHD models of the solar magnetogranulation. The flux density
of the vertical component of the magnetic field (about 200 G) in
the spatial-temporal simulation domain corresponds to the mean
magnetic flux density observed in regions of the photospheric
supergranular network. This suggests that the our series of models
reproduces the region of the supergranular network and its
immediate vicinity in the solar photosphere. Our simulation and
Stokes-diagnostic results qualitatively agree with observations
carried out on the Hinode satellite.

The main conclusions of our analyses are as follows.

(1) On average, the unsigned flux density of the horizontal
magnetic-field component exceeds that of the vertical component at
photospheric heights from 0 to 400 km.

(2) Weak magnetic fields with horizontal components $<$500 G
occupy a larger surface area at the $\tau_5$ = 1 level and make a
larger contribution to the total magnetic energy than do fields
whose vertical component is weaker than 500 G.

(3) Magnetic fields with horizontal components ranging from 500 to
1000 G are concentrated in small islands in the regions between
granules and intergranular lanes. These islands are bipolar in
many cases. They are slightly smaller than the horizontal granular
size, and have lifetimes of 3--6 min.

(4) The fragmentation of a large granule is preceded by the
formation of a strong horizontal tube with field strength
$B_{hor}> 1000$~G at the surface of the granule.

5) The formation of horizontal magnetic fields is closely related
to processes associated with penetrative convection, such as the
expulsion of magnetic flux and the local recirculation of the
granular flux. Their strength depends on the magnetic flux in the
given region.

{\bf Acknowledgments} The author thanks N.V. Kharchenko for
discussions of the results, useful comments and advices.



\end{document}